\documentclass[prl,twocolumn,showpacs,superscriptaddress]{revtex4-1}

\usepackage{amsmath}
\usepackage{bm}
\usepackage{graphicx}

\begin{document}

\title{Slow light in molecular aggregates nanofilms}

\author{E.\ Cabrera-Granado}
\affiliation{Max Planck Institute for the Physics of Complex Systems, 01187 Dresden, Germany}
\affiliation{Universidad Complutense, E-28040 Madrid, Spain}
\author{E.\ D\'{\i}az}
\affiliation{Institute for Materials Science, Technische Universit\"at Dresden, 01062 Dresden, Germany} 
\affiliation{Universidad Complutense, E-28040 Madrid, Spain}
\author{Oscar G.\ Calder\'on}
\affiliation{Universidad Complutense, E-28040 Madrid, Spain}

\begin{abstract}
We study slow light performance of molecular aggregates arranged in nanofilms by means of coherent population oscillations (CPO). 
The molecular cooperative behavior inside the aggregate enhances the delay of
input signals in the GHz range in comparison with other CPO-based devices.
Moreover, the 
problem of residual absorption present in CPO processes, is removed.
We also propose an optical switch between different delays by exploiting the optical bistability of these aggregates. 
\end{abstract}
\maketitle
The optical engineering of the speed of light plays an important role in the development of all-optical devices for telecommunications.
Among the different mechanisms exploited to obtain slow light \cite{Khurgin08, Boyd09}, 
coherent population oscillations (CPO) deals with two-level systems 
and is feasible at room temperature \cite{Gehring08, Bigelow03}. 
It is worth mentioning that there is some controversy in this regard, since most of the experiments can also be explained 
by saturable absorption (SA) \cite{Zapasskii09, Selden09}. Being the source spectral width larger than the coherent 
hole of the absorption profile, some reasonable doubts raise 
about the existence of {\it coherent} population oscillations.
From a theoretical point of view, a rate-equation analysis does not distinguish between
both processes, and a density matrix formalism is required.
To develop photonic applications, a large fractional delay (time delay normalized to the pulse length) 
is desirable in compact devices. 
Furthermore, bandwidths up to the GHz or THz-range and a constrained distortion of the input pulses 
are needed to integrate slow light 
mechanisms in nowadays communication networks. However, the delay decreases with the signal bandwidth, 
which makes difficult to achieve these objectives. Moreover, the residual absorption 
present in CPO processes leads to a greatly diminished output intensity when longer delays are sought. 
In this work we propose a new optical 
CPO-based device considering nanofilms of linear aggregates of dye molecules, so-called J-aggregates, which leads to a fractional delay
up to 0.33 for 135-ps-long pulses while the ratio between the standard deviation of the output 
and input pulses reaches a value of 2. More interestingly, the signal modulation is amplified instead of absorbed, 
which establishes a clear signature to disthinguish between CPO and SA processes in typical two-level systems, since in the latter 
no gain in the weak intensity modulation can be achieved. 
We also show how the already predicted optical bistability on J-aggregates \cite{Malyshev00u} can 
provide an all-optical switch between two well differentiated time delays and pulse distortions. 
Remarkably J-aggregates display narrow absorption bands red-shifted with respect to those of the isolated molecules
due to collective interaction, which leads to interesting optical phenomena studied in the last decades \cite{Knoester06}.
Because of disorder effects affecting these systems,
only some of the molecules of the aggregate are coherently bound.
In the most favorable situation, namely, at low temperatures, the number of molecules over which the excitation is localized is $\sim$100, 
although a real aggregate is consisting of thousands of individual molecules.
In view of these properties, we consider J-aggregates as modeled by an ensemble of inhomogeneously 
broadened two-level systems at low temperatures \cite{Malyshev00u, Malyshev00t}. 
With regard to CPO processes, we study the response of an ultrathin film of oriented linear J-aggregates 
to a strong pump field $E_0$ at frequency $\omega$, and two sidebands
$E_{\pm 1}$ at frequencies $\omega \pm \delta$. Here $\delta$ is the beat frequency between fields. Due to 
disorder effects, the film can be considered as consisting of homogeneous aggregates of different
coherent sizes $N$.
By using the density-matrix formalism under the rotating wave and slowly varying amplitude in time approximations, 
we describe the state of a segment of size $N$ as:
\begin{eqnarray}
\dot{\sigma}^N &=& (i(\omega-\omega_{ba}^N)-1/T_2)\sigma^N-i d^N E Z^N/\hbar\ \nonumber\\
\dot{Z}^N &=& i 2 d^N \left(\sigma^{N*} E - \sigma^N E^*\right)/\hbar-(Z^N + 1)/T_1^N 
\label{densitymatrix}
\end{eqnarray}
Here $\sigma^N$ is the
slowly varying in time coherence between the two levels which depends on the segment size $N$. 
The transition frequency and the dipole moment between the ground (a) and upper (b) energy levels of every segment 
read $\omega_{ba}^N$ and $d^N=d^1\sqrt N$ respectively. The superscript $1$ refers to single-molecule properties. 
We assume field polarization directed along the transition dipole moments of all the aggregates, 
which in addition are parallel to each other as well as to the film plane.
The relaxation time of the population inversion $Z^N=(\rho_{bb}^N-\rho_{aa}^N)$ due to spontaneous emission is 
$T_1^N=T_1^1/N$, while $T_2$ is related to other dephasing processes. 

Field propagation in dense media can be described by means of an integral equation where the slowly varying 
approximation in time is considered but not in space \cite{Malyshev00t}. For film thickness smaller than 
the optical wavelength, spatially homogeneous polarization can be assumed, which leads to the following field equation, 
\begin{equation}
E = E^{in}+\frac{\mu_0 c L}{2} i\omega P \ ,\
\label{eq:fieldgen}
\end{equation}
\noindent where $\mu_0$ is the permeability constant, $c$ is the speed of light and $L$ is the film thickness.
The incident field is $E^{in}(t)=E_0+E_1\exp(-i\delta t)+E_{-1}\exp(i\delta t)$. The second term accounts for 
the field created by the molecules polarized by the incident field.
We checked the full propagation integral to raise the same results as Eq. (\ref{eq:fieldgen}), for
films of tens of nanometers, a thickness achievable by the spin coating technique \cite{Shelkovnikov09}.
Thus, for simplicity and to gain a deeper analytical insight, 
we will restrict ourselves to this case. Equation (\ref{eq:fieldgen}) resembles 
that found by Lu {\it et al.}, who proved that local field effects
can improve the slow light performance in a hybrid nanocrystal complex \cite{Lu08}.

The polarization is calculated by considering the contributions of all coherent segments of different sizes, 
$P=N_0 \sum_N p(N)d^N \sigma^N$. 
Here $N_0$ is the density of localization segments and $p(N)$ refers to the disorder distribution 
function over localization lengths.
Note that the size dispersion of the coherent segments in the system results from the inhomogeneous 
broadening affecting the J-band at low temperatures,
which mainly gives rise to the fluctuation of transition energies $\hbar\omega_{ba}^N$. 
As in Ref. \cite{Malyshev00u}, we replace the average over sizes by one performed over the normalized detunings 
$\Delta^N=(\omega-\omega_{ba}^N)T_2$. 
The distribution $p(N)$ indeed can be accessible by absorption experiments and is considered as Gaussian-like hereafter:
\begin{equation}
\sum_N p(N)\sim\int_{-\infty}^{\infty}\frac{\exp \Big(\frac{-(\Delta^N-\Delta_0)^2}{2 G^2}\Big)}{\sqrt{2 \pi G^2}}d\Delta^N, 
\label{eq:inhom}
\end{equation}
$\Delta_0=(\omega-\omega_0)T_2$ being the detuning between the incident frequency and the mean of the transition
frequency distribution $\omega_0$. The magnitude of the J-bandwidth resulting from the inhomogeneous broadening 
is denoted by $G$ in units of $1/T_2$.
From now on size dispersion effects are restricted to these detuning effects in our calculations, 
for the sake of simplicity.
Thus, we will substitute the size-dependent quantities by its main value in the aggregate and 
remove the index $N$, i. e., $T_1^N=T_1$ and $d^N=d$. 

Similarly to Ref.~\cite{Harter80}, we treat Eqs. (\ref{densitymatrix}) to all orders in the strong field $E_0$, while keeping
only first-order terms in the weak fields $E_{\pm 1}$. 
Within this approximation, the solutions to Eqs.~(\ref{densitymatrix}) are found by considering the Floquet harmonic expansion: 
$ \sigma^N = \sigma_{0}^N + \sigma_{1}^N exp(-i \delta t) + \sigma_{-1}^N exp(i \delta t)$ and $Z^N = Z_{dc}^N + Z_{1}^N exp(-i \delta t) + c.c.$ ~\cite{Harter80}:
\begin{eqnarray}
\sigma_{1}^N &=& \frac{d T_2} {\hbar(\xi+\Delta^N+i)} (E_{1} Z_{dc}^N+E_0 Z_{1}^N) \ ,\nonumber \\
\sigma_{-1}^{N*}&=&\frac{-d T_2} {\hbar(\xi -\Delta^N+i)} (E_0^* Z_{1}^N + E_{-1}^* Z_{dc}^N)\ ,\nonumber \\
Z_{1}^N&=& \frac{-2 d T_2}{\hbar(\xi+i r)} (E_{0}\sigma_{-1}^{N*}+ E_{1} \sigma_{0}^{N*})\nonumber \\
& + &\frac{2 d T_2}{\hbar(\xi+i r)} (E_0^* \sigma_{1}^N + E_{-1}^* \sigma_{0}^N)\ ,
\label{densitymatrixFT}
\end{eqnarray}
where $\xi=\delta T_2$ and $r=T_2/T_1$. The DC response of the population is 
$Z_{dc}$, and $Z_{1}$ 
accounts for the coherent population oscillations, which leads to the absorption dip.

We will refer to the Rabi frequency defined in units of $1/T_2$ as $\Omega_{0,\pm 1}=2dE_{0,\pm 1}T_2/\hbar$ 
from now on as:
\begin{equation}
\Omega_{0,\pm 1}=\Omega_{0,\pm 1}^{in} + i 2 \gamma_{R}\sum_N p(N)\sigma_{0,\pm 1}^N \ .
\label{molfield}
\end{equation}
Here $\gamma_{R}=\mu_0 |d|^2 N_0 c \omega L T_2 / 2\hbar$
is the collective superradiant damping of an ensemble of two-level molecules.

Once we algebraically solve Eqs.~(\ref{densitymatrixFT}) and taking into account Eqs.~(\ref{molfield}) it can be demonstrated that: 
\begin{eqnarray}
\label{ratio1}
\frac{\Omega_0}{\Omega_0^{in}}=  \Big[1+\frac{\gamma_{R}\sum_N p(N)(1 + i\Delta^N)}{1 + (\Delta^{N})^2+|\Omega_0|^2 r^{-1}}\Big]^{-1}\ , 
\end{eqnarray}
\begin{eqnarray}
\label{ratio2}
\frac{\Omega_{1}}{\Omega_{1}^{in}}& =& \Big[1+
i \gamma_R \frac{\Omega_0^2 \Omega_{-1}^*(\xi+2i)}{2\Omega_{1}^{in}} \sum_N \frac{p(N) Z_{dc}^N }{D^{N}(\Delta^N+i)}\Big] \nonumber \\
&\times&\Big[1-i \gamma_R (\xi+i r)\sum_N \frac{p(N)Z_{dc}^N (\xi-\Delta^N+i)}{D^{N}} \nonumber \\
& + & i \gamma_R \frac{ |\Omega_0|^2 \xi }{2} \sum_N \frac{p(N)Z_{dc}^N}{D^{N} (\Delta^N-i)} \Big] ^{-1} \ ,
\end{eqnarray}
\begin{eqnarray}  
\label{ratio3}
\frac{\Omega_{-1}^*}{\Omega_{-1}^{in*}}& =& \Big[1+i \gamma_R \frac{\Omega_0^{*2} \Omega_{1}(\xi+2i)}{2\Omega_{-1}^{in*}} \sum_N \frac{p(N)Z_{dc}^N}{D^{N}(\Delta^N-i)}\Big]\nonumber\\
&\times&\Big[1-i \gamma_R (\xi+i r)\sum_N \frac{p(N)Z_{dc}^N(\xi+\Delta^N+i)}{D^{N}} \nonumber\\
&-& i\gamma_R \frac{ |\Omega_0|^2 \xi}{2} \sum_N \frac{p(N)Z_{dc}^N}{D^{N}(\Delta^N+i)} \Big] ^{-1}\ ,
\end{eqnarray}
where $D^N=(\xi+i r)(\xi+\Delta^N+i)(\xi -\Delta^N+i)-|\Omega_0|^2(\xi+i)$.
Equation~(\ref{ratio1}) has one or three roots depending on the parameters. This leads to bistable solutions 
for the strong field $\Omega_0$ when $\gamma_R$ is larger than a threshold value~\cite{Malyshev00u}.

We consider an incident field such that $\Omega_{1}^{in}=\Omega_{-1}^{in*}$, leading to 
a sinusoidally modulated intensity with frequency $\delta$. 
Thus, Eqs.~(\ref{ratio2}) and~(\ref{ratio3}) can be simplified and the resulting fields 
fulfill $\Omega_{1}=\Omega_{-1}^*$.
The transmittance $T$ and the dephasing $\phi$ induced by the film is calculated by the ratio between the output and the input signals:
\begin{equation}
\frac{\Omega_{1}+\Omega_{-1}}{\Omega_{1}^{in}+\Omega_{-1}^{in}}=T e^{i \phi} ,
\label{eq:fd}
\end{equation}
Thus, the fractional delay is defined as $F=\phi/2\pi$.

We first analyze the case in absence of size dispersion. We take the parameters of Pseudoisocyanine-Br (PIC-Br) 
as it is one of the most studied J-aggregates. 
Hence, we use $T_1 = 37$ ps
(corresponding to a homogeneous aggregate of size $N=100$) and $T_2 = 0.02 T_1$. This value is consistent
with measurements at low temperatures \cite{Fidder90} and allows direct comparison with previous 
CPO works \cite{Boyd81}.
The transition dipole moment is $d^1 = 12.1$ D and the concentration of
aggregates is $N_0 \sim 10^{23}$ m$^{-3}$. These parameters give values for $\gamma_R$ around 10 for film
lengths of tens of nanometers.
As in the usual CPO case, the maximum delay is obtained when the pump field reaches the saturation 
intensity, which depends on $\gamma_R$. Hereafter, and for each case shown, we will restrict ourselves to 
this optimal intensity. 
These values correspond to a photon flux of $\sim 1.4\times 10^{11}$ phot/cm$^2$ for a 100-ps-length pulse,
and a surface density of monomers $\sim 10^{13}$ cm$^{-2}$. This means that the typical number of absorption and re-emission 
cycles per aggregate is $\sim$ 1, a much smaller value than those usually occurring in J-aggregates
experiments \cite{Markov03}, and that allows to ensure photostability of the samples. 
Moreover these intensities are low enough 
to neglect the blue-shifted one-to-two exciton transitions, which supports the use of a two-level model \cite{Malyshev00u}.

\begin{figure}[ht]
\centerline{\includegraphics[width=80mm,clip]{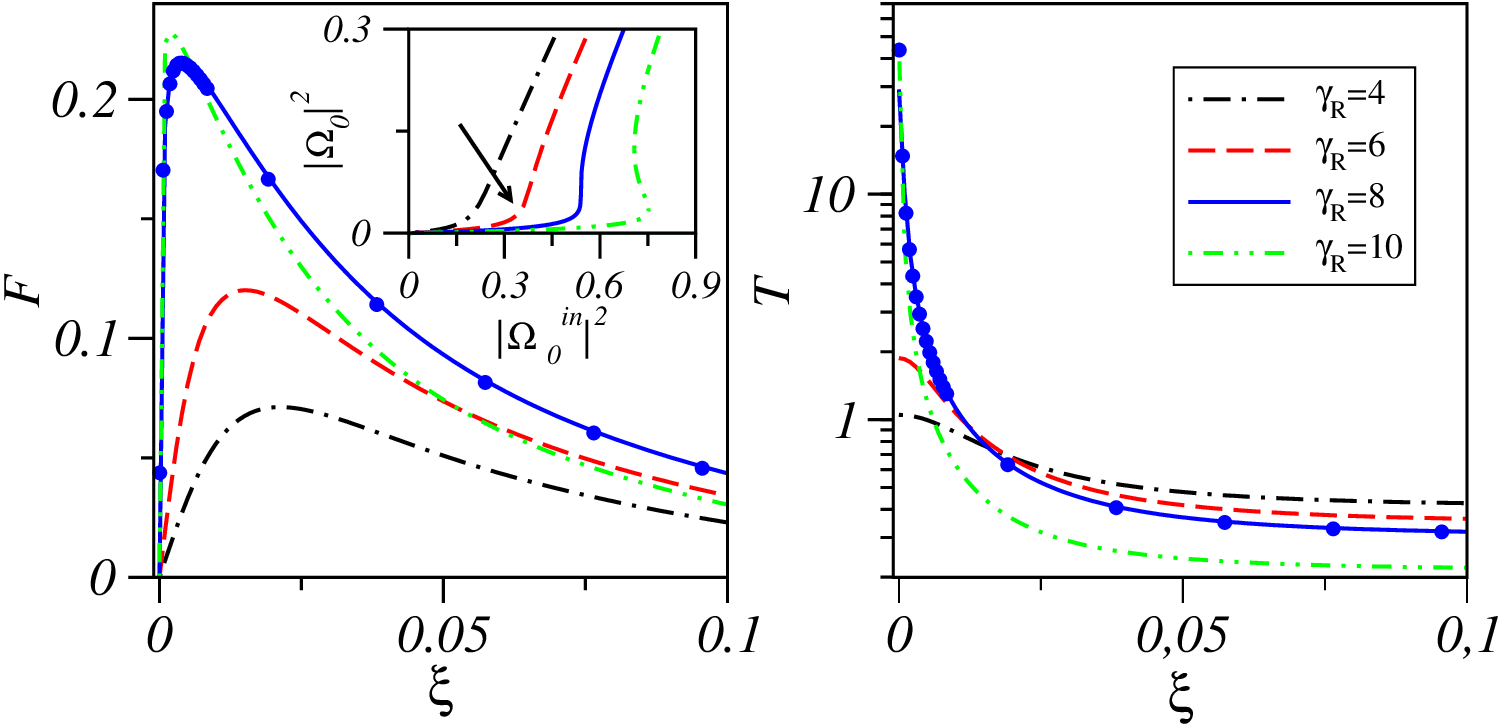}}
\caption{Fractional delay and transmittance of a sinusoidal input signal as a function of the dimensionless 
detuning $\xi=\delta T_2$ for different values of $\gamma_R$ and $G=0$. The inset shows the Output-Input curve for the strong field $\Omega_0^2$. 
The arrow points the optimal intensity for the case $\gamma_R = 6$. Solid dots result from numerical integration of Eqs. (\ref{densitymatrix}) and (\ref{eq:fieldgen}).}
\label{fig1}
\end{figure}

Figure \ref{fig1} shows the fractional delay and transmittance, calculated 
with Eq. (\ref{eq:fd}), as a function of the normalized modulation frequency and for different values 
of $\gamma_R$. An increasing value of this parameter, which accounts for the collective interaction within
the aggregate, gives rise to higher fractional delays reaching values of $\sim$0.2 for 
$\gamma_R \sim 8$. Values of $\gamma_R > 8$ allows bistability to appear, which introduces 
a large signal distortion. 
Therefore, only
cases up to $\gamma_R = 10$ are shown. Figure \ref{fig1} (right panel) shows how the transmission 
increases with  $\gamma_R$ as well. It must be noted that the system presents gain ($T > 1$) in the weak probe 
electric fields
for modulation frequencies close to those revealing the maximum fractional delay. We explain this behaviour as a
strong scattering process from the pump to the weak field due to the transient temporal grating generated in CPO. 
This is clearly relevant as 
CPO processes in other media usually present non-desirable residual absorption. Hence, in such cases it may be 
necessary to amplify the signal after propagation through the slow light medium, specially for long media or 
high ion-densities \cite{Melle07}. The presence of gain in these films could overcome this 
drawback of CPO-based slow light while is revealed as a clear difference with SA processes in typical two-level media.

\begin{figure}[ht]
\centerline{\includegraphics[width=80mm,clip]{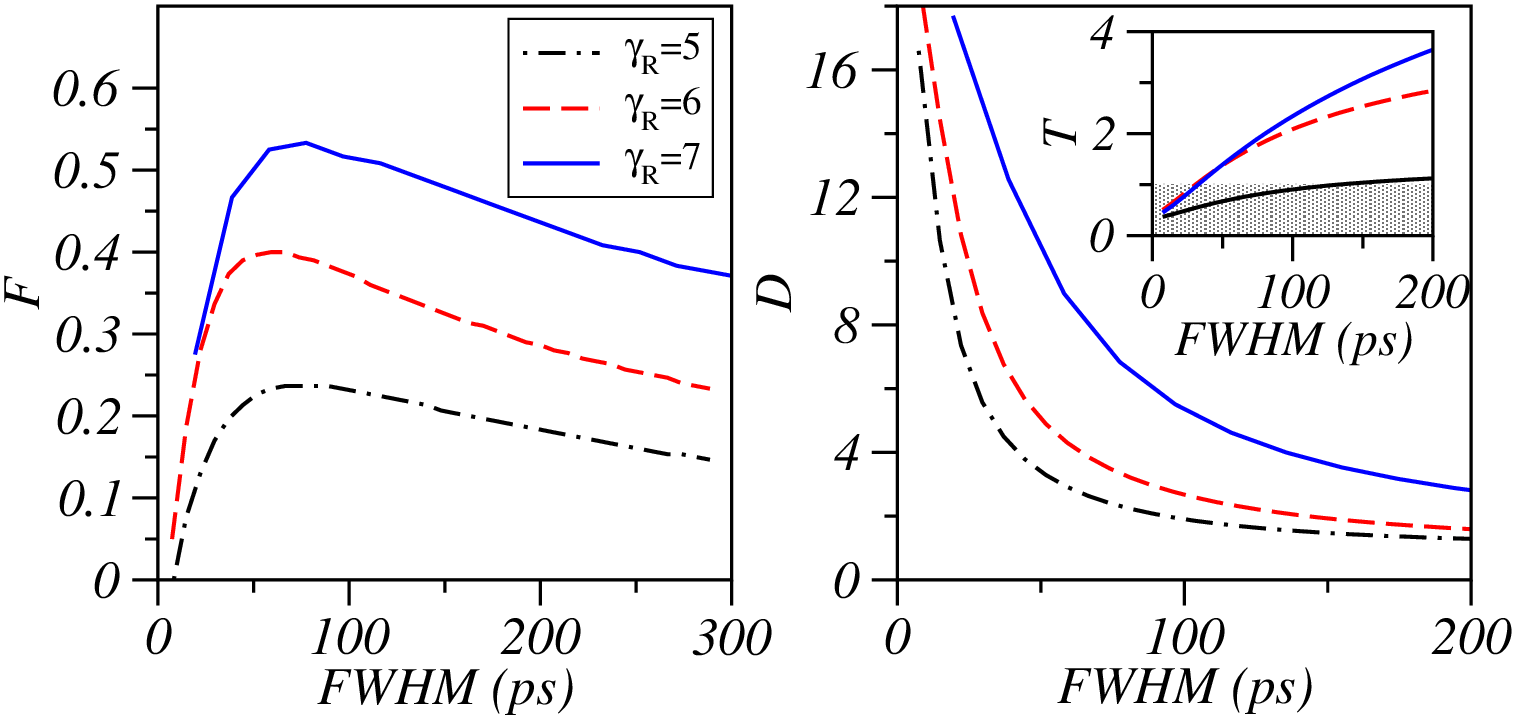}}
\caption{Fractional delay and distortion for pulsed input signals against the initial pulse temporal width.
Different values of $\gamma_R$ are considered. Inset: Transmission vs FWHM.}
\label{fig2}
\end{figure}

More relevant to telecommunication applications is the performance of the system with input pulse signals. To analyze the fractional delay and distortion 
of the output pulses after passing through the ultrathin film
we numerically integrate Eqs. (\ref{densitymatrix}) together with Eq. (\ref{eq:fieldgen}). 
To validate the simulations, we first check the numerical integration with
the results previously obtained. Figure \ref{fig1} shows the perfect agreement with the analytical predictions for 
the case of a sinusoidal input signal. Figure \ref{fig2} depicts the fractional delay ($F$)
and distortion ($D$) as a function of the pulse temporal width (defined as FWHM). The distortion is defined as the 
ratio between the output and input-pulse standard 
deviations. Although the maximum fractional delay is accompanied
by a large distortion, values up to $F = 0.33$ are obtained with distortion 2 for 135-ps-long pulses, 
which gives $\sim$7.5-GHz-bandwidth. These results notably improve previous data obtained by CPO-based slow 
light in semiconductor materials at GHz-bandwidths \cite{Palinginis05}.
As mentioned before, increasing values of 
$\gamma_R$ lead to larger delays. However, the distortion generated when bistability occurs limits in 
practice the values of $\gamma_R$ up to 8 for G = 0 (no size dispersion). Note in Fig. \ref{fig2} (inset)
how the transmittance of the electric field is larger than 1 for most of the analyzed input-pulse widths, 
in agreement with results of Fig. \ref{fig1}.

\begin{figure}[ht]
\centerline{\includegraphics[width=80mm,clip]{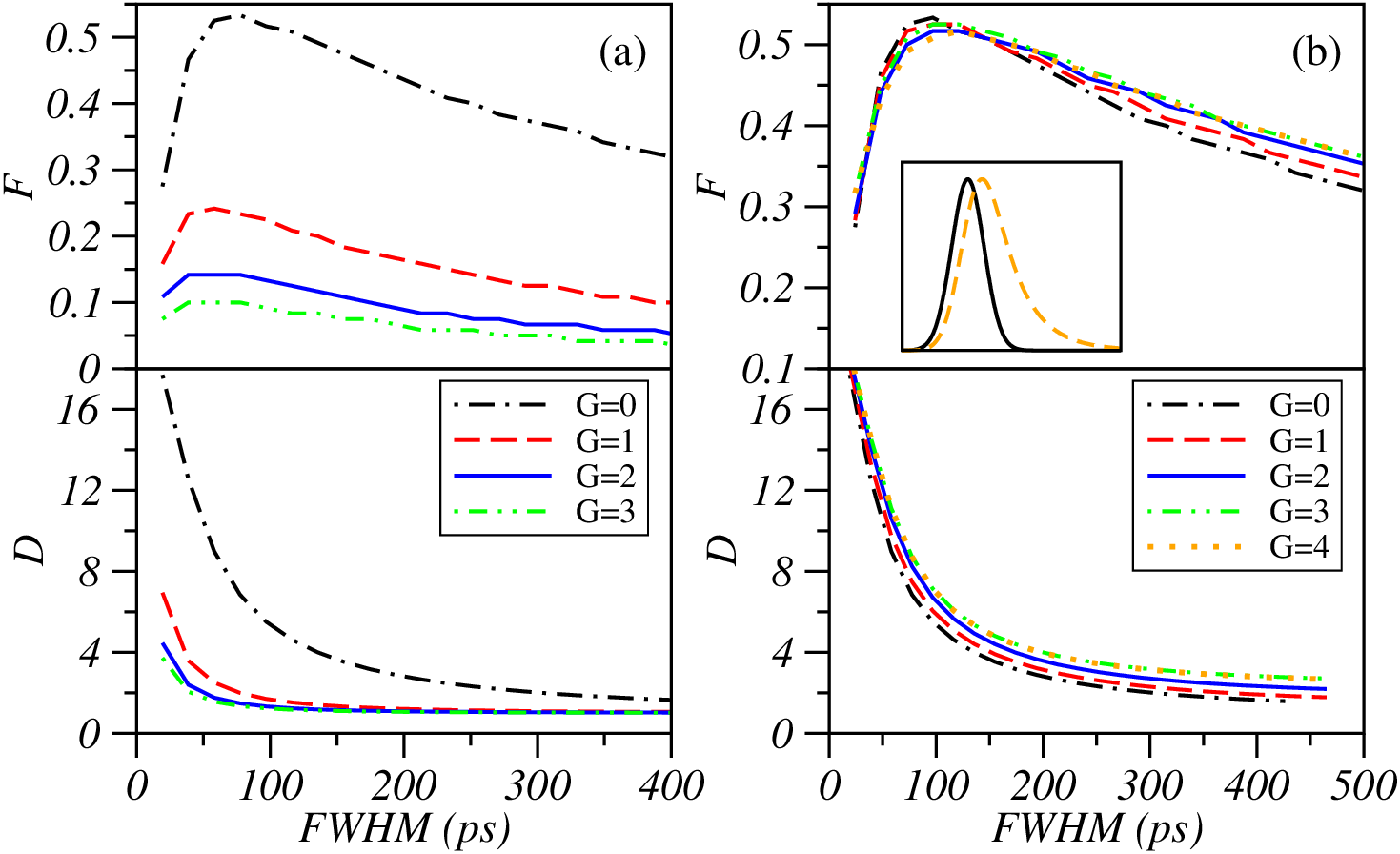}}
\caption{Fractional delay and distortion for input pulses vs the initial pulse temporal width (FWHM), 
for different values of $G$ and (a) for $\gamma_R = 7$, (b) for $\gamma_R = 7$ (G=0), $\gamma_R = 12$ (G=1), 
$\gamma_R = 22$ (G=2),$\gamma_R = 34$ (G=3),$\gamma_R = 49$ (G=4).}
\label{fig3}
\end{figure}

Let us now analyze the influence of the size dispersion on the slow light performance. To numerically 
integrate Eqs. (\ref{densitymatrix}) and (\ref{eq:fieldgen}) 
with the inclusion of the inhomogeneous broadening, we carefully choose
the sampling under the curve defined by Eq. (\ref{eq:inhom}) to reproduce the analytical 
results for a sinusoidal signal. We now focus on the effects of size dispersion on pulse propagation.
As it is well known, the presence of inhomogeneous broadening reduces the slow light performance, as can 
be seen in Fig. \ref{fig3}, where the fractional delay and distortion are plotted against the input-pulse width. 
A value of $G = 3$ already reduces the fractional delay up to 5 times with respect to the value obtained without
size dispersion. However, as shown at right panels of Fig. \ref{fig3}, the detrimental effect of a larger 
inhomogeneous linewidth can be compensated by increasing the 
value of $\gamma_R$. 
Larger values of $\gamma_R$ can be obtained by modifying the temperature or increasing the aggregates concentration in the sample \cite{Malyshev00u}.

\begin{figure}[ht]
\centerline{\includegraphics[width=80mm,clip]{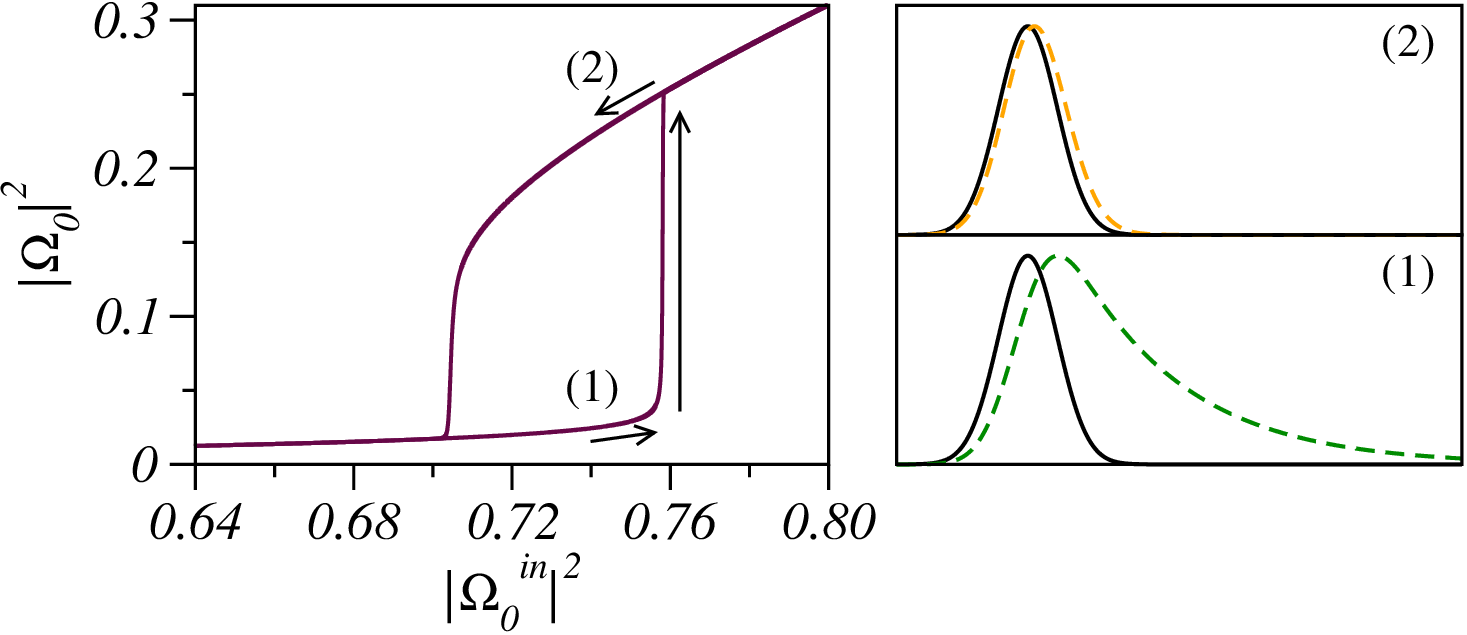}}
\caption{Switch between different delays through the bistability loop for a pulse FWHM = 111 ps. 
Arrows mark the change of $|\Omega_0^2|$ when applying a short pulse in the input signal. In position (1) $F =  0.43$ 
while $F = 0.09$  in (2). Output and input pulses  in (1) and (2) are shown at the right panel.}
\label{fig4}
\end{figure}

We finally propose a mechanism to take advantage of the optical bistability present in J-aggregates nanofilms for 
slow light applications.
Bistability allows to rapidly change between two different output intensities for the same input. As the fractional
delay depends on $|\Omega_0|^2$, this property turns out into a switch between two well-differentiated delays
and distortions for the same $|\Omega_0^{in}|^2$. Fig \ref{fig4} depicts this process for a value of $\gamma_R = 10$ and no
size dispersion, where the bistability loop can be found for $0.69 \le (\Omega_0^{in}) ^2 \le 0.76$. Starting in
the lower branch of the loop (position (1) in Fig \ref{fig4}), and after propagating the first pulse, 
a short pulse in the input signal causes the system
to switch to position (2), in the upper branch. Fractional delays achieved in positions (1) and (2), for a 111-ps-length 
pulse are 0.43 and 0.09 respectively, while the distortion values are 3.6 and 1.07. The lowest switching time 
is limited by $T_1$, acording to simulations, being of some tens of ps for J-aggregates films.

In conclusion we showed that J-aggregates ultrathin films can produce large fractional delays by means of CPO processes,
even in presence of inhomogeneous broadening and thanks to the cooperative behavior of the aggregate molecules. This system
does not suffer from residual absorption in the weak probe fields, in opposition to the usual CPO-based slow light. We also
demonstrated how optical bistability could be used to produce a fast switch between different delays. We believe these
organic compounds present a viable alternative to semiconductor slow light devices in the nanoscale. To this end, our results
can motivate further research at room temperature operation and telecommunication bands. In this sense,
the development of new aggregates such as Porpho-cyanines \cite{Hales09} provides great opportunities. Moreover, new studies including                
processes such as one-to-two exciton transitions and exciton-exciton anhilition are currently in progress. These effects
seems to counteract the killing action of the inhomogeneous broadening \cite{Glaeske01}, which would impose better experimental conditions  
to obtain the delays shown in this work.
\begin{acknowledgments}
This work was supported by projects MOSAICO, MAT2010-17180 and CCG08-UCM/ESP-4330. E. D. 
acknowledges financial support 
by Ministerio de Eduacion y Ciencia. 
We thank A. Eisfeld for
valuable discussions.
\end{acknowledgments}

\end{document}